\documentstyle[eqsecnum,aps,epsf]{revtex} 

\begin{document}

\title{A new {\it ab initio} method of calculating $Z_{eff}$ and  
positron annihilation rates using coupled-channel T-matrix amplitudes}

\author{P K Biswas}
\address{
Departamento de F\'{\i}sica, Instituto Tecnol\'ogico de Aeron\'autica,
CTA\\
S\~ao Jos\'e dos Campos 12228-901, SP, Brasil\\
email: biswas@fis.ita.br}

\maketitle

\begin{abstract}

A new {\it ab initio} theoretical formulation to calculate $Z_{eff}$ and
hence the positron annihilation rates is presented using the on-shell and
half-offshell T-matrix scattering amplitudes without any explicit use of
the scattering wave function.  The formulation is exact and universal, as
long as the dynamics of the systems can be described by a
Lippmann-Schwinger type equation. It could serve as an effective tool as
all the $T-$, $K-$, and $S-$matrix formulations, yield directly the
scattering amplitudes; not the wave function. 
We also present a basic numerical test of the formulation.

{\bf PACS Number(s):  3.65.Nk, 34.85.+x, 71.60.+z, 78.70.Bj}

\end{abstract}

\newpage

\section{Introduction}
Positron, being an antiparticle, interacts intimately (short-range
interaction) with atomic electrons due to the absence of any restriction
imposed by the Pauli Exclusion principle. Thus their annihilation studies,
namely, the {\it Positron annihilation spectroscopy} and {\it Positronium
annihilation lifetime spectroscopy} have emerged as two front line
research areas, as they are expected to provide a more detail account of
their close interactions with the target and bear the potential of various
modern technological applications \cite{pws,appl1,appl2}. 


Up until now, the theoretical study of annihilations rate requires the 
evaluation
of the scattering wave function.  However, it is of fundamental importance
to note that most scattering calculations (T-, K-, S-matrix) yield
directly the scattering amplitudes; not the wave function. Thus, the
evaluation of $Z_{eff}$ normally requires a separate calculation for the
wave function using these scattering amplitudes (like those in 
ref.\cite{ft1}), or need to adopt of a different methodology that will 
yield the scattering wave function directly.  Ryzhikh and Mitroy 
\cite{ft1} have
used the T-matrix formulation to evaluate the scattering wave function
which has subsequently been used to evaluate $Z_{eff}$. Gribakin \cite{gg}
has used an approximate form of the T-matrix equation to evaluate
$Z_{eff}$. Here, we present a new {\it ab-initio} theoretical 
formulation based on the integral equation
formalism, whereby $Z_{eff}$ and hence the annihilation rates can be
exactly calculated using on-shell and half-offshell T-matrix scattering
amplitudes {\it but without any explicit use of the scattering wave
function}.

Theoretically, the positron annihilation 
rates ($\Lambda$)  are expressed in terms of $Z_{eff}$,
the {\it effective}
number of target electrons available to the incoming positron as
\cite{humb,map1,map2}:
\begin{eqnarray} \label{e1}
\Lambda &=& \pi r_0^2cZ_{eff}N \mbox{ sec}^-1
\end{eqnarray}
And $Z_{eff}$ is defined in terms of the 
scattering wave function $|\psi_k^+\rangle$ as \cite{humb,map1,map2}
\begin{eqnarray}\label{e2}
Z_{eff}({\bf k}) &=& \langle\psi_{k}^+
|\sum_{j=1}^N\delta({\bf r}_j-{\bf x})|\psi_{k}^+\rangle
\end{eqnarray}
where $r_0$ is the classical electron radius; $c$ is the speed of light;
$4\pi r_0^2c$ is the non-relativistic electron-positron 
annihilation rate; $\pi r_0^2c$ is the same for the spin-averaged case of 
two-gamma annihilation (which excludes electron-positron triplet state 
contribution and considers only singlet state annihilation).
$N$ is the number density of atoms or molecules in the medium;
$\delta$ is the Dirac $\delta$-function;
${\bf x}$ and ${\bf r}_j$ are the positron and the electron 
co-ordinates.

Here we present a new formulation whereby the {\it r.h.s} of equation
\ref{e2} is represented exactly by the onshell and half-offshell
coupled-channel T-matrix scattering amplitudes, thus facilitating the
evaluation of $Z_{eff}$ without any explicit use of the scattering wave
function. When a positron collides with a target, it can have direct as
well as rearrangement scattering. So, we present our formulation in two
different sections consisting of {\bf I}) direct (elastic and inelastic)
scattering and {\bf II}) direct plus rearrangement (positronium formation)  
scattering.

\section{Positron Annihilation Considering the Direct Scattering 
Channels}

In this section, we  consider the direct 
scattering of positrons from atomic ($\phi_n$) targets where the
total wave function is expanded as \cite{greview}:
\begin{eqnarray}\label{e3}
\psi_k^+({\bf x},{\bf r}_1,{\bf r}_2,...{\bf r}_N)&=& \sum_n
F_n({\bf x})\phi_n ({\bf r}_1,{\bf r}_2,...{\bf r}_N)
\end{eqnarray}
where $F_n$'s are the expansion coefficients, representing the  
motion of the positron with 
momentum ${\bf k}$; ${\bf r}_j$ is the co-ordinate of the $j-th$ 
electron and ${\bf x}$ is that of the positron. 
The total Hamiltonian is partitioned as 
\begin{eqnarray}
H=H_d^0+V_d
\end{eqnarray}
where $H_d^0$ is the unperturbed part of the total 
Hamiltonian in the direct scattering channel of the positron and the atom 
and $V_d$ is the 
positron-atom interaction potential. The unperturbed and the total 
Hamiltonians satisfy the following eigen-value equations:
\begin{eqnarray}
H_d^0|k\phi_n\rangle &=& E|k\phi_n\rangle \label{e4} \\
(H_d^0+V_d)|\psi_k^+\rangle &=& E|\psi_k^+\rangle \label{e5}
\end{eqnarray}
where $E=k^2/2m-{\cal E}_A$ is the total energy; ${\cal E}_A$ 
is the binding energy of the initial target atom ($\phi_n$); $m$ and 
${\bf k}$ 
are the reduced mass and the onshell momentum of the positron.
With the eigen-value equations \ref{e4} and \ref{e5} for the unperturbed 
and the total Hamiltonians, one can write the Lippmann-Schwinger 
equation for the scattering wave function $|\psi_k^+\rangle$ as 
\cite{greview,glockle}:
\begin{eqnarray}\label{e6}
|\psi^+_k\rangle=|k\phi_n\rangle +\frac{1}{E-H_d^0+i0}V_d|\psi^+_k\rangle
\end{eqnarray}
Using the closure relation 
$(2\pi)^{-3}\sum_{n''}\int d^3k''
|k''\phi_{n''}\rangle\langle k''\phi_{n''}|=1$, 
and using the T-matrix definition: 
$V_d|\psi^+_k\rangle=T|k\phi_n\rangle$, in eqn({\ref{e6}),
we arrive at the 
expression for the total scattering wave function in terms of the 
$T$-matrix elements:
\begin{eqnarray}\label{e7}
|\psi^+_k\rangle=|k\phi_n\rangle + 
\frac{1}{(2\pi)^3}\sum_{n''=1}^\infty\int d^3k'' 
\frac{|k''\phi_{n''}\rangle\langle k''\phi_{n''}|T|k\phi_n\rangle}
{E-E_n''+i0}
\end{eqnarray}
The total scattering wave function can be evaluated from this equation 
\cite{ft1}.
However, we generally solve it for the scattering T-matrix amplitudes 
obtained by multiplying 
eqn.(\ref{e7}) by $V_d$ and projecting with $\langle k'\phi_{n'}|$
and using the T-matrix definition: $V_d|\psi^+_k\rangle=T|k\phi_n\rangle$:
\begin{eqnarray}\label{et}
\langle k'\phi_{n'}|T|k\phi_n\rangle &=&
\langle k'\phi_{n'}|V_d|k\phi_n\rangle 
\nonumber \\
&+& \frac{1}{(2\pi)^3}\sum_{n''}\int d^3k'' 
\frac{\langle k'\phi_{n'}|V_d
|k''\phi_{n''}\rangle\langle k''\phi_{n''}|T|k\phi_n\rangle}
{E-E_n''+i0}
\end{eqnarray}
Eqn.(\ref{et}), in its one-dimensional partial-wave form (eqn.\ref{tj}), 
is 
exactly solved using the matrix inversion method 
\cite{gh1}. Simultaneous equations are formed by replacing $k'$ with 
various values of $k''$ on which the radial integral over $dk''$ is 
discretized. The solutions of the simultaneous equations give us the 
both-onshell 
($\langle k'\phi_{n'}|T|k\phi_n\rangle$) 
and half-offshell 
($\langle k''\phi_{n'}|T|k\phi_n\rangle$) 
$T$-matrix amplitudes for various values of $k''$, where ${\bf k,k'}$ are 
on shell momenta and 
$\bf{k''}$ are the off-shell ones. 
While the solutions for the onshell $T$-matrix elements reflect the 
asymptotic behavior of the wave function and provide 
the physical cross 
sections, the half-offshell elements are usually thrown away. 
We understand 
that the latter might contain the short-range properties of the wave 
function and they 
together with the on-shell elements can lead to an exact evaluation of 
the $Z_{eff}$. 
We multiply eqn.{\ref{e7} from left by $\sum_{j=1}^N\delta({\bf x}-{\bf 
r}_j)=\Delta, \mbox{ (say)}$ and 
project it by $\langle\psi_k^+|$ to obtain:
\begin{eqnarray}\label{e9}
\langle \psi^+_k|\Delta|\psi^+_k\rangle &=&
\langle\psi^+_k|\Delta|k\phi_n\rangle 
\nonumber \\
&+& \frac{1}{(2\pi)^3}\sum_{n''}\int d^3k'' 
\frac{\langle \psi^+_k|\Delta
|k''\phi_{n''}\rangle\langle k''\phi_{n''}|T|k\phi_n\rangle}
{E-E_n''+i0}
\end{eqnarray}
At this stage, to calculate $Z_{eff}$, which is equivalent to 
$\langle\psi_k^+|\Delta|\psi_k^+\rangle$ (see eqn.\ref{e2}), we have 
two options: 1) using equation \ref{e7}, substitute for 
$\langle\psi_k^+|$ in the $r.h.s$ of (\ref{e9}) 
or 2) evaluate 
$\langle\psi_k^+|\Delta|k\phi_n\rangle$ separately and substitute in 
eqn(\ref{e9}). The first case leads to a complicated equation as follows: 
\begin{eqnarray}\label{ezz}
\langle \psi^+_k|\Delta|\psi^+_k\rangle &=&
\langle k\phi_n|\Delta|k\phi_n\rangle 
\nonumber \\
&+& \frac{1}{(2\pi)^3}\sum_{n''}\int d^3k'' 
\frac{\langle k\phi_n|T|k''\phi_{n''}\rangle
\langle k''\phi_{n''}|\Delta|k\phi_n\rangle}
{E-E_n''-i0}\nonumber \\
&+& \frac{1}{(2\pi)^3}\sum_{n''}\int d^3k'' 
\frac{\langle k\phi_n|\Delta
|k''\phi_{n''}\rangle\langle k''\phi_{n''}|T|k\phi_n\rangle}
{E-E_n''+i0} \nonumber \\
&+& \frac{1}{(2\pi)^6}\sum_{n''}\sum_{m''}\int d^3k_1''\int d^3k_2'' 
\frac{\langle k\phi_n|T|k_1''\phi_{n''}\rangle
\langle k_1''\phi_{n''}|\Delta|k_2''\phi_m''\rangle
\langle k_2''\phi_{m''}|T|k\phi_n\rangle}
{(E-E_n''-i0)(E-E_m''+i0)} \nonumber\\
\end{eqnarray}
This equation, although can be solved numerically, needs an extra effort 
to evaluate the principal value part of the last term, which contains 
a product of two singularities arising out of the product of Greens 
functions. We, therefore, look for the evaluation 
of $\langle k\phi_n|\Delta|\psi_k^+\rangle$ by projecting eqn(\ref{e7}) 
with $\langle 
k\phi_n|\Delta$:  
\begin{eqnarray}\label{e8}
\langle k'\phi_{n'}|\Delta|\psi^+_k\rangle &=&
\langle k'\phi_{n'}|\Delta|k\phi_n\rangle 
\nonumber \\
&+& \frac{1}{(2\pi)^3}\sum_{n''}\int d^3k'' 
\frac{\langle k'\phi_{n'}|\Delta
|k''\phi_{n''}\rangle\langle k''\phi_{n''}|T|k\phi_n\rangle}
{E-E_n''+i0}
\end{eqnarray}
We 
solve this equation exactly (which is a very straight forward 
numerical summation) 
and 
substitute the complex conjugate of $\langle 
k'\phi_{n'}|\Delta|\psi_k^+\rangle$ in 
eqn(\ref{e9}) to get $Z_{eff}$. However, like the T-matrix equation, 
we solve them in their one-dimensional partial wave form. To 
arrive at the corresponding 
partial wave equations for (\ref{e8}) and (\ref{e9}),
we define the matrices $D$ and $Z$ as:
\begin{eqnarray}
\Delta|\psi_k^+\rangle &= & D|k\phi_n\rangle \\
\langle\psi_k^+|D &=& \langle k\phi_n|Z
\end{eqnarray}
and rewrite eqn(\ref{e8})  and 
eqn(\ref{e9}) formally in terms of them:
\begin{eqnarray}\label{e8a}
\langle k'\phi_{n'}|D|k\phi_n\rangle &=&
\langle k'\phi_{n'}|\Delta|k\phi_n\rangle 
\nonumber \\
&+& \frac{1}{(2\pi)^3}\sum_{n''}\int d^3k'' 
\frac{\langle k'\phi_{n'}|\Delta
|k''\phi_{n''}\rangle\langle k''\phi_{n''}|T|k\phi_n\rangle}
{E-E_n''+i0}
\end{eqnarray}
and
\begin{eqnarray}\label{e9a}
\langle k\phi_n|Z|k\phi_n\rangle &=&
\langle k\phi_n|\bar{D}|k\phi_n\rangle 
\nonumber \\
&+& \frac{1}{(2\pi)^3}\sum_{n''}\int d^3k'' 
\frac{\langle k\phi_n|\bar{D}
|k''\phi_{n''}\rangle\langle k''\phi_{n''}|T|k\phi_n\rangle}
{E-E_n''+i0}
\end{eqnarray}
where $\bar{D}$ is the complex conjugate of $D$.
Using a partial wave decomposition of the form:
\begin{eqnarray}
\langle{ k}'\phi_{n'}|X|{ k}\phi_n\rangle &=&
\sum_J\sum_M\sum_L\sum_{M_L}\sum_{L'}\sum_{M_{L'}}
\langle L'l',M_{L'}m_{l'}|JM\rangle Y^*_{L'M_{L'}}(\hat{\bf k'})
\nonumber\\
&&\langle Ll,M_{L}m_{l}|JM\rangle Y_{LM_{L}}(\hat{\bf k})
X_{J} ( {n'l'L'k',nlLk})
\end{eqnarray}
where, $X\equiv T,V,D,\Delta,\hskip 2pt \mbox{or} \hskip 2pt Z$;
$n,l$ are the 
principal and orbital quantum number of the target and $L$ 
is the orbital quantum number of the moving positron in the initial 
state; primed quantities denote the same for the final state.
With the above expansion, the 
scattering T-matrix 
equation and the above two equations 
for the $D$- and $Z$-matrices reduce to:
\begin{eqnarray} \label{tj} 
T_J(\tau ',k'; \tau,k)&=& V_J( \tau ',k';\tau k)
\nonumber \\
&+&\frac{m''}{4\pi^3} \sum_{\tau{''}} \int dk{''}{k''}^2\frac
{V_J(\tau ',k';\tau{''},k'')
T_J(\tau{''},k'';\tau,k)}
{k_{\tau{''}}^2-k{''}^2+i0} 
\end{eqnarray} 
\begin{eqnarray} \label{dj} 
D_J(\tau{'},k'; \tau,k)&=& \Delta_J( \tau{'},k';\tau k)
\nonumber \\
&+&\frac{m''}{4\pi^3} \sum_{\tau{''}} \int d{k''}{k''}^2\frac
{\Delta_J(\tau{'},k';\tau{''},k'')
T_J(\tau{''},k'';\tau,k)}
{k_{\tau{''}}^2-k{''}^2+i0} 
\end{eqnarray} 
\begin{eqnarray} \label{zj} 
Z_J(\tau{'},k'; \tau,k)&=& \bar{D}_J( \tau{'},k';\tau k)
\nonumber \\
&+&\frac{m''}{4\pi^3} \sum_{\tau{''}} \int d{k''}{k''}^2\frac
{\bar{D}_J(\tau{'},k';\tau{''},k'')
T_J(\tau{''},k'';\tau,k)}
{k_{\tau{''}}^2-k{''}^2+i0} 
\end{eqnarray} 
where $\tau\equiv (nlL)$ and $\tau'\equiv (n'l'L')$; $m''$ is the 
reduced mass of the projectile in the intermediate state (here, $m''=m=1$ 
in au). We suppress the suffix $d$ from $V_d$ for convenience.

In terms of partial wave $Z$-matrices, $Z_{eff}(k^2)$ comes out to be: 
\begin{eqnarray}
Z_{eff}(k^2)=\sum_J \frac{2J+1}{4\pi} 
Z_J(nlLk;nlLk)
\end{eqnarray}

While eqn.(\ref{tj}) is generally used to study positron-atom scattering, 
eqs.(\ref{dj}) and (\ref{zj}) are particularly useful to evaluate 
$Z_{eff}$ from the onshell and half-offshell T-matrix outputs of 
eqn.\ref{tj}). We shall present a simple numerical account on $e^+$-He 
scattering 
to verify the code and compare the numbers, but beforehand we 
present a general formula for $Z_{eff}$ by inserting eqn.(\ref{dj}) into 
eqn.(\ref{zj}). This latter is of particular interest, as it will 
explicitly demonstrate how  the $Z_{eff}$ is dependent on the T-matrices. 
For this, we first 
rewrite eqs.(\ref{tj}), (\ref{dj}) and (\ref{zj}) in 
the following notations: 
\begin{eqnarray}
T_{k'k}&=&V_{k'k}-iV_{k'k'}T_{k'k}
+V_{k'k''}G_0(k',k'')T_{k''k} 
\label{dd0}\\
D_{k'k}&=&\Delta_{k'k}-i\Delta_{k'k'}T_{k'k}
+\Delta_{k'p''}G_0(k',p'')T_{p''k} 
\label{dd1}\\
Z_{k'k}&=&\bar{D}_{k'k}-i \bar{D}_{k'k'}T_{kk}
+\bar{D}_{k'q''}G_0(k',q'')T_{q''k} 
\label{dd2}
\end{eqnarray} 
where summations over intermediate states are implied and off-shell 
momenta are represented by $k''$, $p''$ and 
$q''$. In the above,
we have used the following relation for the complex Greens function 
\begin{eqnarray}\label{g0}
G_0^+(k^2-{k''}^2)=\frac{1}{k^2-{k''}^2+i0}=
-i\pi\delta(k^2-{k''}^2)+\frac{P}{k^2-{k''}^2}
\end{eqnarray} 
to expand it into real and imaginary parts; $P$ represents principal value 
integration; $G_0$ represents the real (principal value) part of $G_0^+$.
Inserting $\bar{D}_{kk}$ in eqn(\ref{dd2}) we obtain an explicit relation
for $Z_{eff}(k^2)\equiv Z_{kk}$:
\begin{eqnarray}
Z_{kk}&=& \Delta_{kk}
+i\Delta_{kk}[T_{kk}^*-T_{kk}]
+T_{kp''}G_0\Delta_{p''k}
+\Delta_{kq''}G_0T_{q''k}\nonumber \\
&+&\Delta_{kk}|T_{kk}|^2
+T_{kp''}G_0\Delta_{p''q''}G_0T_{q''k} \label{dff1}\\
&=& \Delta_{kk}
+2\Delta_{kk}\mbox{Im}[T_{kk}]
+\Delta_{kk}|T_{kk}|^2
+T_{kp''}G_0\Delta_{p''k}
+\Delta_{kq''}G_0T_{q''k}\nonumber\\
&+&T_{kp''}G_0\Delta_{p''q''}G_0T_{q''k} \label{dff2}
\end{eqnarray}
where $\Delta_{pq}$ corresponds to plane-wave value of $Z_{eff}$ for 
the initial 
and final momenta $p$ and $q$. Im$[T_{kk}]$ and $|T_{kk}|^2$ are
proportional to physical cross sections (representing 
the asymptotic behavior of the wave function). Others are interference 
terms, linear and quadratic in the half-offshell T-matrix 
elements, and 
expected to play a crucial role at low and intermediate energies.
To understand their role, and to check the normalizations of 
equations \ref{dj} and \ref{zj},
we provide a numerical test below.

\subsection{Numerical test to equations \ref{dj}, \ref{zj}}
To test the formulations of the equations \ref{dj}, \ref{zj}}
(Equation \ref{tj} is well established in literature) we perform sample 
calculation on $e^+$-He elastic scattering, 
considering only the static 
interaction [keeping summation over 
$\tau''\equiv$ He$(1s^2)$], and evaluate the $Z_{eff}$ using the 
resulting onshell 
and half-offshell T-matrix elements.  We use atomic units 
throughout and use delta-function normalization for the plane wave.

We note that equations \{\ref{tj},\ref{dj},\ref{zj}\} and 
\{\ref{dd0},\ref{dd1},\ref{dd2}\} are equivalent. 
In figure 1, first we plot the dotted curve which is obtained 
considering only the plane wave parts (first term of the $r.h.s$) 
of eqs.(\ref{dd0}), (\ref{dd1}) and (\ref{dd2}).
This plane wave approximation gives a value of $Z_{eff}=2.0 (=Z)$ as was 
expected and provides the normalization. 
Next we consider first two terms of the $r.h.s$ of
eqs.(\ref{dd0}), (\ref{dd1}) and (\ref{dd2}) and plot the result as 
dashed curve. This approximation is equivalent of considering the plane 
wave and the onshell T-matrix contributions of equations (\ref{tj}),
(\ref{dj}) and (\ref{zj}) leaving aside the 
half-offshell contribution (particular integral part of the Greens 
function). We obtain a 
lower value of $Z_{eff}$ with increasing energies, signaling the 
manifestation of a repulsive potential at higher energies. Now, the solid 
curve is obtained with all 
the three terms  of eqs.(\ref{dd0}), (\ref{dd1}), and (\ref{dd2}). That 
is, considering both the onshell and the half-offshell contributions 
together with the plane wave. The static 
potential in a $e^+$-He scattering is repulsive and consequently it 
lowers the value of $Z_{eff}$ (and hence the annihilation rate).
We compare the solid curve with the results of a Schwinger Multi-Channel 
(SMC) calculation on $e^+$-He with the same physical content (considering 
static 
interaction only) \cite{pvt}. Both the curves agree quite well. 
The marginal 
difference in the $Z_{eff}$ value between the present and the SMC 
calculation is supposed to be acceptable since,
the wave functions for Helium used in these two calculations 
are different (we use the Roothaan-Hatree-Fock five-term wave function 
of Clementi and Roetti \cite{rhf} for He) and also the T-matrix and the 
SMC formulations are different.

The results provided in figure-1 are aimed at checking the 
new equations vis-a-vis other methodologies where we considered the static 
interaction. Formally, to
arrive at a physically converged result for the $Z_{eff}$, full expansion 
basis indicated in equations \ref{dj} and \ref{zj} need to be employed 
like the case for the scattering T-matrix equation \ref{tj}.
However, in practice it has been found that the positron-scattering 
cross sections do not converge
 easily unless real and virtual effects of the rearrangement 
channel of positronium (Ps) formation is considered in the 
theoretical formulation. So, for a converged description of the scattering 
and annihilation we need to consider the Ps formation channel explicitly 
in the theoretical formulation. However, the above formulation is suitable 
for employing model polarization potentials alongside the static potential 
so as to arrive at a meaningful physically converged result without 
being confined in the {\it{ab initio}} framework.

\section{Positron Annihilation Considering the Direct and the Ps-formation 
Channels.}

When the possibility of a real or virtual positronium formation is
considered, through the capture of a target electron by the incident
positron, the theoretical formulation for a single electron target differs
from a multi electron target in the sense that for the latter case the
Ps-target(ion) wave function need to be formally antisymmetrized.  
Here, we discuss them in two different sections {\bf A} and {\bf B}.

\subsection{Single Electron Target}
For positron scattering from a single electron 
target, the total wave function (\ref{e3}) can be expanded (considering Ps 
formation) as \cite{greview,wl}:
\begin{eqnarray}\label{2e1}
\psi_k^+({\bf x},{\bf r}_1)&=& \sum_n
F_n({\bf x})\phi_n ({\bf r}_1) 
+\sum_{\nu}
{\cal G}_{\nu}({\bf \rho}_1)\chi_\nu({\bf t}_1)
\end{eqnarray}
where ${\bf {\rho}}_1=({\bf r}_1+{\bf x})/2$ and ${\bf t}_1={\bf r}_1-{\bf 
x}$. ${\cal G}_{\nu}$ and $\chi_\nu$ 
represent the moving and the bound-state
($\nu$-th) positronium atom.
The total Hamiltonian is now partitioned as: 
\begin{eqnarray}\label{part}
H=H_d^0+V_d=H_{c}^0+V_{c}
\end{eqnarray} 
where $H_d^0$, $H_{c}^0$ are the unperturbed  
Hamiltonians in the direct ($d$) and capture ($c$) channels 
satisfying the eigen-value equations
\begin{eqnarray}   
H_d^0|k\phi_{n}\rangle &=& E_n|k\phi_{n}\rangle \label{ev1}\\
H_c^0|k\chi_{\nu}\rangle &=& E_\nu|k\chi_{\nu}\rangle \label{ev2}
\end{eqnarray} 
and $V_d$ and $V_c$ are the interaction potentials therein. 
$E_n=k_x^2/2-{\cal E}_A$ and $E_\nu=k_{Ps}^2/4-{\cal E}_{Ps}$;
 ${\cal E}_A$ and ${\cal 
E}_{Ps}$ are the binding energies of the initial target atom and the 
rearranged positronium atom; $k_x$ and $k_{Ps}$ are the momenta of the 
positron and the positronium. 
In terms of the two-cluster channel-Greens-functions 
$G_d^0=(E-H_d^0)^{-1}$ and  $G_c^0=(E-H_c^0)^{-1}$, 
we take the Lippmann-Schwinger integral 
equation for the wave function as \cite{fadeev}:
\begin{eqnarray}\label{2e2}
|\psi_k^+\rangle = |k\phi_n\rangle + G_d^0T_d|k\phi_n\rangle
+G_c^0T_c|k\phi_n\rangle
\end{eqnarray} where,
$T_d$ and $T_c$ are defined 
as $V_d|\psi^+_k\rangle=T_d|k\phi_n\rangle$  
(here $T_d\equiv T$ of section-I) 
and
$V_c|\psi^+_k\rangle=T_c|k\phi_n\rangle$. 
Using the following
closure relations for the direct and the rearrangement channels,  
\begin{eqnarray}
1&=& \frac{1}{(2\pi)^3}\sum_{n''}\int dk'' 
|k''\phi_{n''}\rangle\langle k''\phi_{n''}|\\
1&=& \frac{1}{(2\pi)^3}\sum_{\nu''}\int dk'' 
|k''\chi_{\nu''}\rangle\langle k''\chi_{\nu''}|
\end{eqnarray}
we rewrite eqn.(\ref{2e2}) as:
\begin{eqnarray}
|\psi^+_k\rangle &=&|k\phi_n\rangle 
+ \frac{1}{(2\pi)^3}\sum_{n''}\int d^3k'' 
\frac{|k''\phi_{n''}\rangle\langle k''\phi_{n''}|T_d|k\phi_n\rangle}
{E-E_n''+i0} \nonumber\\
&+& \frac{1}{(2\pi)^3}\sum_{\nu''}\int d^3k'' 
\frac{|k''\chi_{\nu''}\rangle\langle 
k''\chi_{\nu''}|T_c|k\phi_n\rangle}
{E-E_\nu''+i0} \label{2e3}
\end{eqnarray}
Here,
$E_n''={k''}^2/2-{\cal E}_A$ and $E_\nu''={k''}^2/4-{\cal E}_{Ps}$ are 
the off-shell energies 
in the direct (d) and the capture (c) channels.
We construct the coupled equations by 
1) multiplying this equation with 
$V_d$ and projecting out with $\langle k'\phi_{n'}|$
 and
2) multiplying this equation with 
$V_c$ and projecting out with $\langle k'\chi_{\nu'}|$
\begin{eqnarray}\label{2ez1}
\langle k'\phi_{n'}|T_d|k\phi_n\rangle &=&
\langle k'\phi_{n'}|V_d|k\phi_n\rangle 
+ \frac{1}{(2\pi)^3}\sum_{n''}\int d^3k'' 
\frac{\langle k'\phi_{n'}|V_d
|k''\phi_{n''}\rangle\langle k''\phi_{n''}|T_d|k\phi_n\rangle}
{E-E_n''+i0} \nonumber\\
&+& \frac{1}{(2\pi)^3}\sum_{\nu''}\int d^3k'' 
\frac{\langle k'\phi_{n'}|(E_\nu''+V_c-E_{n'})|
k''\chi_{\nu''}\rangle\langle 
k''\chi_{\nu''}|T_c|k\phi_n\rangle}
{E-E_\nu''+i0}
\end{eqnarray}
\begin{eqnarray}\label{2ez2}
\langle k'\chi_{\nu'}|T_c|k\phi_n\rangle &=&
\langle k'\chi_{\nu'}|V_c|k\phi_n\rangle 
+ \frac{1}{(2\pi)^3}\sum_{n''}\int d^3k'' 
\frac{\langle k'\chi_{\nu'}|V_c
|k''\phi_{n''}\rangle\langle k''\phi_{n''}|T_d|k\phi_n\rangle}
{E-E_n''+i0} \nonumber\\
&+& \frac{1}{(2\pi)^3}\sum_{\nu''}\int d^3k'' 
\frac{\langle k'\chi_{\nu'}|V_c|
k''\chi_{\nu''}\rangle\langle 
k''\chi_{\nu''}|T_c|k\phi_n\rangle}
{E-E_\nu''+i0}
\end{eqnarray}
where in eqn.(\ref{2ez1}), we use $V_d=H_c^0+V_c-H_d^0$ (see 
eqn.\ref{part}) and also use the eigen-value equations 
(\ref{ev1},\ref{ev2}).
Once the above coupled-equations are solved and we are equipped with the 
T-matrix 
amplitudes 
$\langle p\phi_{n'}|T_d|k\phi_n\rangle$ and 
$\langle q\chi_{\nu'}|T_c|k\phi_n\rangle$ for on-shell and 
off-shell values for the momenta $p$ and $q$, we can get $Z_{eff}$ in 
terms of them.  To deduce $Z_{eff}$, in terms of T-matrices, we project 
equation (\ref{2e3}) 
by $\langle\psi_k^+|\Delta$ and arrive at: 
\begin{eqnarray}\label{33}
\langle\psi_k^+|\Delta|\psi^+_k\rangle &=&
\langle\psi_k^+|\Delta|k\phi_n\rangle 
+ \frac{1}{(2\pi)^3}\sum_{n''}\int d^3k'' 
\frac{\langle\psi_k^+|\Delta
|k''\phi_{n''}\rangle\langle k''\phi_{n''}|T_d|k\phi_n\rangle}
{E-E_n''+i0} \nonumber\\
&+& \frac{1}{(2\pi)^3}\sum_{\nu''}\int d^3k'' 
\frac{\langle\psi_k^+|\Delta |k''\chi_{\nu''}\rangle
\langle k''\chi_{\nu''}|T_c|k\phi_n\rangle}
{E-E_\nu''+i0}
\end{eqnarray}
Now, $\langle\psi_k^+|$ from eqn(\ref{2e3}) may be substituted in the 
$r.h.s$ of 
eqn(\ref{33}) to arrive at a direct expression for $Z_{eff}$. However, 
that will lead to a complicated equation like (\ref{ezz}). 
We rather develop simpler equations to evaluate 
$\langle \psi_k^+|\Delta|k\phi_n\rangle$ and
$\langle \psi_k^+|\Delta|k\chi_\nu\rangle$ and substitute them back in 
eqn(\ref{33}). For this, we
project eqn(\ref{2e3}) from left by 
$\langle k'\phi_{n'}|\Delta$ and 
$\langle k'\chi_{\nu'}|\Delta$ and obtain:
\begin{eqnarray}\label{34}
\langle k'\phi_{n'}|\Delta|\psi^+_k\rangle &=&
\langle k'\phi_{n'}|\Delta|k\phi_n\rangle 
+ \frac{1}{(2\pi)^3}\sum_{n''}\int d^3k'' 
\frac{\langle k'\phi_{n'}|\Delta
|k''\phi_{n''}\rangle\langle k''\phi_{n''}|T_d|k\phi_n\rangle}
{E-E_n''+i0} \nonumber\\
&+& \frac{1}{(2\pi)^3}\sum_{\nu''}\int d^3k'' 
\frac{\langle k'\phi_{n'}|\Delta |
k''\chi_{\nu''}\rangle\langle 
k''\chi_{\nu''}|T_c|k\phi_n\rangle}
{E-E_\nu''+i0}
\end{eqnarray}
\begin{eqnarray}\label{35}
\langle k'\chi_{\nu'}|\Delta|\psi^+_k\rangle &=&
\langle k'\chi_{\nu'}|\Delta|k\phi_n\rangle 
+ \frac{1}{(2\pi)^3}\sum_{n''}\int d^3k'' 
\frac{\langle k'\chi_{\nu'}|\Delta
|k''\phi_{n''}\rangle\langle k''\phi_{n''}|T_d|k\phi_n\rangle}
{E-E_n''+i0} \nonumber \\
&+& \frac{1}{(2\pi)^3}\sum_{\nu''}\int d^3k'' 
\frac{\langle k'\chi_{\nu'}|\Delta |k''\chi_{\nu''}\rangle\langle 
k''\chi_{\nu''}|T_c|k\phi_n\rangle}
{E-E_\nu''+i0}
\end{eqnarray}
The above two equations are very straight forward 
to solve as one need to carry only numerical integrations with
known values of $T_d$, $T_c$ and the calculated plane-wave matrix 
elements concerning $\Delta$ as inputs. 
We are not interested to repeat the calculations for 
$T_d$ and $T_c$ and 
rather hope that the existing T-matrix results \cite{ppp} may be applied 
to calculate $Z_{eff}$.

\subsection{Many Electron Target}
For multi-electron targets the formulation is very much similar to 
that of section IIA, except few 
fundamental changes. Without repeating the whole thing, we thus mention 
here 
about the necessary changes. For the positron scattering from a 
multi-electron target, the 
capture channel need to be explicitly antisymmetrized and expressed as:
\begin{eqnarray}\label{3e1}
\psi_k^+({\bf x},{\bf r}_1,{\bf r}_2,..{\bf r}_N)&=& \sum_n
F_n({\bf x})\phi_n ({\bf r}_1,....,{\bf r}_N) 
+{\cal A}_1\sum_{\nu\mu}
{\cal G}_{\nu\mu}({\bf \rho}_1)\chi_\nu({\bf t}_1)\varphi_\mu({\bf 
r}_2,...,{\bf 
r}_N)
\end{eqnarray}
where $\varphi$ represents the residual target ion and ${\cal A}_1$ is the 
antisymmetrization operator, which antisymmetrizes electron 1 with other 
target electrons. The initial target wave function $\phi$ is supposed to
be antisymmetrized implicitly.
The total Hamiltonian is now partitioned as: 
$H=H_d^0+V_d=H_{c(j)}^0+V_{c(j)}$; 
where $H_{c(j)}^0$ and $V_{c(j)}$ are the unperturbed  
Hamiltonian and the Ps-target(ion) interaction potential in the 
capture channel of the positronium formation, with the $j$-th electron 
being attached to the positron. Accommodating the Pauli exclusion 
principle for the rearrangement channel, the 
Lippmann-Schwinger integral equation is now written as:
\begin{eqnarray}\label{3e2}
|\psi_k^+\rangle = |k\phi_n\rangle + G_d^0T_d|k\phi_n\rangle
+{\cal A}_jG_{c(j)}^0T_{c(j)}|k\phi_n\rangle
\end{eqnarray}
where $T_d$ and $T_c$ are defined 
as $V_d|\psi^+_k\rangle=T_d|k\phi_n\rangle$  
(here $T_d\equiv T$ of section-I) 
and
$V_{c(j)}|\psi^+_k\rangle=T_{c(j)}|k\phi_n\rangle$. 
Using following closure relations for the direct and the rearrangement 
channels:
\begin{eqnarray}
1&=& \frac{1}{(2\pi)^3}\sum_{n''=1}^\alpha\int dk'' 
|k''\phi_{n''}\rangle\langle k''\phi_{n''}|\\
1&=& \frac{1}{(2\pi)^3}\sum_{\nu''}\sum_{\mu''}\int dk'' 
|k_j''\chi_{\nu''}\varphi_{\mu''}\rangle
\langle k_j''\chi_{\nu''}\varphi_{\mu''}|
\end{eqnarray}
and proceeding in a similar way, we represent the Lippmann-Schwinger 
equation \ref{3e2} as:
\begin{eqnarray}
|\psi^+_k\rangle &=&|k\phi_n\rangle 
+ \frac{1}{(2\pi)^3}\sum_{n''}\int d^3k'' 
\frac{|k''\phi_{n''}\rangle\langle k''\phi_{n''}|T_d|k\phi_n\rangle}
{E-E_n''+i0} \nonumber\\
&+& \frac{1}{(2\pi)^3}\sum_{\nu''}\sum_{\mu''}\int d^3k'' 
\frac{{\cal A}_j|k_j''\chi_{\nu''}\varphi_{\mu''}\rangle\langle 
k_j''\chi_{\nu''}\varphi_{\mu''}|T_{c(j)}|k\phi_n\rangle}
{E-E_{\nu\mu}''+i0} \label{3e3}
\end{eqnarray}
The rest of the procedures are exactly similar to those described in 
section IIA and are not repeated here.

In summary, we present a new {\it ab initio} methodology to calculate
$Z_{eff}$ from physical (onshell) and virtual (half-offshell) scattering
T-matrix amplitudes, without any use of the scattering wave function.  
The formulation presented here is for positron annihilation in atoms, but
it could be universally applied to other annihilation studies as long as
the dynamics of the interacting particles (or clusters)  can be described
by the well-known Lippmann-Schwinger type equation.  The methodology is
exact and can act as an useful tool for the annihilation studies as most
of the scattering theories (T-matrix, K-matrix, S-matrix) yield directly
the scattering amplitudes. Performing elastic scattering and employing
elastic-channel T-matrix amplitudes (on- and off-shell) we reproduce the
corresponding $Z_{eff}$ result and demonstrate the utility of the 
methodology. A similar 
T-matrix
formulation for the pick-off annihilation of ortho-positronium is under
way and will be published soon.

The work has been carried out under financial support from FAPESP, Brazil
through project number 99/06844-7. I gratefully acknowledge various
discussions with Dr. T. Frederico, Dr. J. S. E.  Germano of our department
and Dr. M.  A. P. Lima and Mr. M. Varella of UNICAMP, SP, Brazil.

\newpage
{\bf Figure Caption:}

Figure 1. Theoretical values of $Z_{eff}$ in various approximations as a 
function of positron energy for the target of atomic helium.


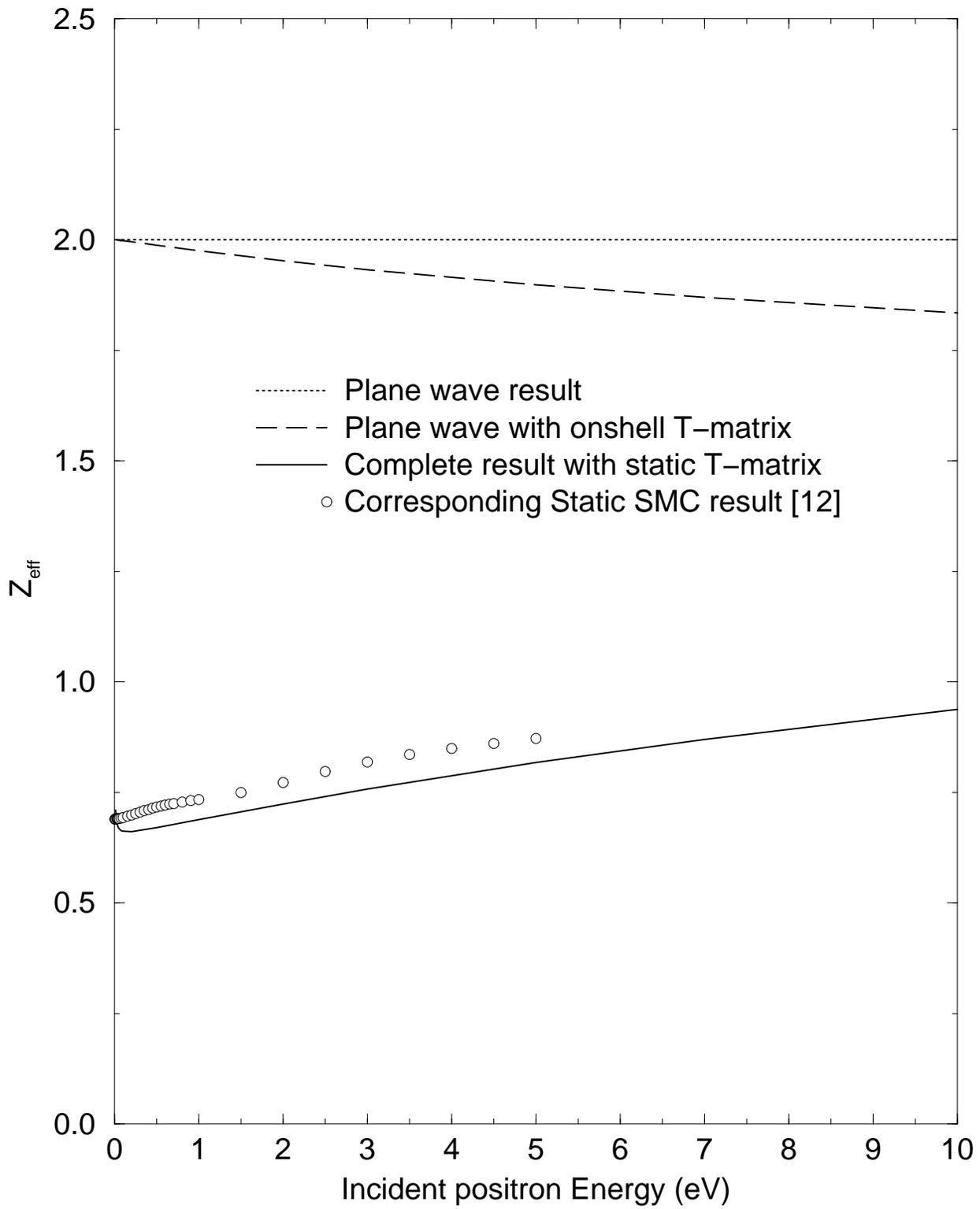
\end{document}